\documentclass[%
reprint,
superscriptaddress,
 amsmath,amssymb,
 aps,
 pra,
longbibliography,
dvipsnames,
twocolumn,
xcolor=table
]{revtex4-1}
\usepackage[main=english,german,ngerman]{babel}
\usepackage{graphicx}
\usepackage{dcolumn}
\usepackage{bm}
\usepackage{siunitx}
\usepackage{dsfont}

\usepackage{mathtools }
\usepackage{subcaption}
\usepackage{hyperref}
\usepackage[table,xcdraw]{xcolor}

\usepackage{physics}
\usepackage{bbold}
\usepackage{float}
\usepackage{hyphenat}
\newcommand\figref{Fig.~\ref}
\usepackage[utf8]{inputenc}
\usepackage[toc,page]{appendix}

\usepackage{graphicx}
\usepackage{hyperref}
\hypersetup{colorlinks}
\usepackage{soul}

\begin{document}
\newcommand{\orcidicon}[1]{\href{https://orcid.org/#1}{\includegraphics[height=\fontcharht\font`\B]{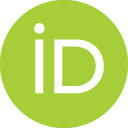}}}
\title{Conditional Born machine for Monte Carlo event generation}

\author{Oriel~Kiss\orcidicon{0000-0001-7461-3342}}
\email{oriel.kiss@cern.ch}
\affiliation{European Organization for Nuclear Research (CERN), Geneva 1211, Switzerland}
\affiliation{Department of Computer Science, University of Geneva, Geneva 1211, Switzerland}

\author{Michele~Grossi\orcidicon{0000-0003-1718-1314}}
\affiliation{European Organization for Nuclear Research (CERN), Geneva 1211, Switzerland}

\author{Enrique~Kajomovitz}
\affiliation{Departement of Physics, Technion, Haifa 32000, Israel}

\author{Sofia~Vallecorsa\orcidicon{0000-0002-7003-5765}}
\affiliation{European Organization for Nuclear Research (CERN), Geneva 1211, Switzerland}

\renewcommand*\labelenumi{(\theenumi)}

\date{\today}

\begin{abstract}

Generative modeling is a promising task for near\hyp term quantum devices, which can use the stochastic nature of quantum measurements as a random source. So called Born machines are purely quantum models and promise to generate probability distributions in a quantum way, inaccessible to classical computers. This paper presents an application of Born machines to Monte Carlo simulations and extends their reach to multivariate and conditional distributions. Models are run on (noisy) simulators and IBM Quantum superconducting quantum hardware.

More specifically, Born machines are used to generate muonic force carrier (MFC) events resulting from scattering processes between muons and the detector material in high\hyp energy physics colliders experiments. MFCs are bosons appearing in beyond\hyp the\hyp standard\hyp model theoretical frameworks, which are candidates for dark matter. Empirical evidence suggests that Born machines can reproduce the marginal distributions and correlations of data sets from Monte Carlo simulations.
\end{abstract}

\maketitle

\section{Introduction}
Quantum computers have the potential to solve problems that are difficult for classical computers, such as factoring \cite{Shor} or simulation of quantum systems \cite{QPE}. However, the unavailability of error\hyp correcting codes and limited qubit connectivity prevents them from being used. Nevertheless, noisy\hyp intermediate\hyp scale\hyp quantum (NISQ) \cite{Preskill2018quantumcomputingin} devices, characterized by their low number of noisy qubits and short decoherence time, have already been proven to be successful in domains such as machine learning \cite{QML_Lloyd,variational_quantum_algorithm,QCL, QNN_FF,Schuld_PRL,QCNN, feature_map,QGAN_Zoufal,Born_machine_Coyle_2021,Born_machine_Liu, Intro_QML} and quantum chemistry  \cite{VQE_Gambetta, UCC-chemistry, magnetogrossi}.

The present paper focuses on generative modeling in quantum machine learning (QML), which is the task of learning the underlying probability distribution $\pi(y)$ of a given data set and generating samples from it. In the classical regime, generative models are often expressed as neural networks. For instance, generative adversarial networks (GANs) \cite{GAN_Goodfellow} and variational autoencoders \cite{VAE} have been successfully applied in a variety of fields, ranging from computer vision \cite{CV_GAN} to natural sciences \cite{ML_science}. In high\hyp energy physics (HEP), generative models have been proposed as an alternative to Monte Carlo (MC) simulations, e.g., to simulate detectors \cite{GHEP, CaloGAN, Maevskiy_2020} and, very recently, as a method to load distributions of elementary particle-physics processes \cite{agliardi2022quantum}. MC calculations in HEP, such as GEANT4 \cite{GEANT4:2002zbu} and MADGRAPH \cite{Mad-graph}, are usually expensive in time and CPU resources \cite{cpu2018}. Generative models provide a solution, e.g., by augmenting small MC data sets or interpolating or extrapolating to different regimes.

The probabilistic nature of quantum mechanics allows us to define a new class of generative models: \textit{quantum circuit Born machines} (QCBMs). These models use the stochastic nature of quantum measurement as randomlike sources and have no classical analog. More specifically, they produce samples from the underlying distribution of a pure quantum state by measuring a parametrized quantum circuit \cite{PQC}  with probability given by the Born rule $p_\theta(x)=|\bra{x}\ket{\psi(\theta)}|^2$. Born machines have been proposed as Bayesian models \cite{Born_machine_Liu}, using an adversarial training strategy \cite{QGAN_Zoufal,QGAN_Chang}, using optimal transport \cite{Born_machine_Coyle_2021}, in a conditional setting \cite{Cond_Born}, and adapted to continuous data \cite{Born_machine_CV,GAN_continous}. Quantum neural networks using a Gaussian noise source \cite{style_qgan} and quantum Boltzmann machines \cite{Boltzmann_machine_Zoufal} are both viable alternatives for quantum generative modeling but will not be addressed in the present paper. Quantum generative models also have the ability to load probability distribution on a quantum computer \cite{QGAN_Zoufal}, which can then be used to integrate elementary processes via quantum amplitude estimation \cite{agliardi2022quantum}, for finance applications \cite{Born_machine_finance}, or for variational inference \cite{Cond_Born} tasks.

Here, an extension to multivariate and conditional probability distributions is proposed, exploring the limitations of NISQ devices. Generating multivariate distributions with Born machines was already explored in \cite{Joint_distribution,QCBM_MC}, and the latter reference also focuses on HEP applications. Moreover, we introduce a conditional Born machine for the generation of probability distributions depending on an external parameter. The contributions of the present paper are as follows:
\begin{enumerate}
    \item An alternative circuit design with a reduced connectivity is proposed, which is better suited for NISQ devices. \\
    \item While \cite{Cond_Born} focuses on conditional distributions in a Bayesian setting, we instead propose a Born machine conditioned on a parameter of the MC simulations, which can be used to efficiently generate data in different regimes. \\
    \item All simulations are mapped to real quantum devices and executed.
\end{enumerate}

Experiments were conducted on (noisy) simulators and superconducting devices from IBM Quantum (IBMQ) using QISKIT\hyp RUNTIME, the recent serverless architecture framework which handles classical and quantum computations simultaneously on a dedicated cloud instance. Noisy simulators incorporate gates and readout errors, approaching real device performances. However, their behavior can be genuinely different from noisy simulators. Hence, this work emphasizes the use of real quantum hardware and addresses related challenges.

This paper is organized as follows. Section \ref{MFC} introduces the physical\hyp use case of muonic force carriers and the preprocessing of the data set. The models are introduced in Sec. \ref{model}. More specifically, Sec. \ref{QCBM} introduce the quantum circuit Born machine, and Secs. \ref{3DBM} and \ref{CBM} the multivariate and conditional versions, respectively. Section \ref{training} explains the training strategy. Results are shown in Sec. \ref{res} for all models on (noisy) simulators and real quantum hardware and are discussed in Sec. \ref{discussion}.

\section{Muonic force carriers}
\label{MFC}
\subsection{Physical setting}
Muon force carriers (MFCs) are theorized particles that could be constituents of dark matter and explain some anomalies in the measurement of the proton radius and the muon's magnetic dipole, making them exciting candidates for new physics searches.

Following \textcite{MFC_Kajomovitz}, we consider a muon fixed\hyp target scattering experiment between muons produced at the high\hyp energy collisions of the Large Hadron Collider and the detector material of the Forward Search Experiment (FASER) or the ATLAS experiment calorimeter. In the ATLAS case \cite{MFC_Kajomovitz}, independent muon measurements performed by the inner detector and muon system can help us observe new force carriers coupled to muons, which are usually not detected. In the FASER experiment, the high\hyp resolution of the tungsten\hyp emulsion detector is used to measure the muon's trajectories and energies. 
\subsection{Data set}
The dataset \cite{MFC_dataset}, produced with MADGRAPH5 simulations \cite{Mad-graph}, is composed of samples with the following variables: the energy $E$, transverse momentum $p_t$ and pseudorapidity $\eta$ of the outgoing muon and  MFC, conditioned on the energy of the incoming muon. The data are made more Gaussian shaped by being preprocessed in the following way: the energy is divided by the mean of the incoming energy, the transverse momentum is elevated to the power of 0.1 \cite{power_transform}, and everything is standardized to zero mean and unit variance. The purpose of the preprocessing is to ease the training, and the preprocessing has shown improvement over the generation of the same events using classical GAN \cite{MFC_GAN}. The dataset is composed of 10 240 distinct events, and it is split into training and testing sets of equal size.

\section{Models}
\label{model}
\subsection{Quantum circuit Born machine}
\label{QCBM}
\begin{figure}
    \centering
    \includegraphics[scale =0.5]{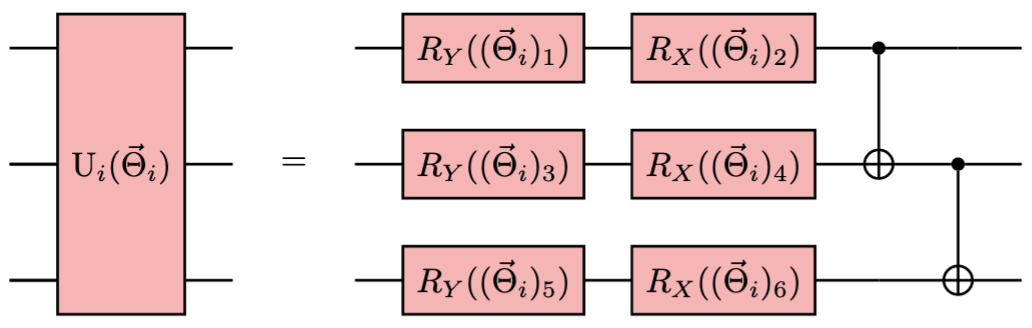} 
    \caption{Ansatz constructed with $R_y(\theta)$ and $R_x(\theta)$ single\hyp qubit rotation and CNOT interactions with a linear connectivity.}
    \label{fig:U_ha}
\end{figure}

A Born machine represents a probability distribution as a quantum pure state and can generate samples via projective measurements. The Born machine outputs binary strings, which can be interpreted as a sample from the generated discrete probability distribution. Similar to a classification task, the target distribution is discretized into $2^N$ bins, which are associated with different binary strings of size $N$. The quantum state can take the form of a quantum circuit \cite{Born_machine_Liu} or a tensor network \cite{TNBM}, acting on some initial state, e.g., $\ket{0}^{\otimes N}$. \textcite{QCBM_MC} numerically demonstrated that the initial state has only a negligible impact on the training, reason why this simple, physic-independent state is chosen. This paper considers quantum circuit Born machines, where the quantum circuit $U(\Theta)$ is constructed, for convenience, using $L$ repetitions of basic layers $U_i(\vec{\Theta}_i)$, where $\Theta$ = $\{\vec{\Theta}_0,...,\vec{\Theta}_L\}$ is the set of all parameters. $\vec{\Theta}_i$ is a vector of trainable parameters for the specific layer $i$ which needs to be trained to match all amplitudes for the desired $N$-qubit registers to find the corresponding state.  These building blocks are chosen as a hardware\hyp efficient ansatz \cite{VQE_Gambetta}, which can be run on current quantum chips with minimal overhead. An example constructed with $R_y(\theta) =  \exp(-i\theta \sigma_y/2)$ and $R_x(\theta) = \exp(-i\theta \sigma_x/2)$ single\hyp qubit rotations,  where $\theta \in [0,2\pi)$, are trainable parameters, and controlled NOT (CNOT) interaction between two qubits with linear connectivity is shown in \figref{fig:U_ha}. Here, $\sigma_k$ ($k=x,y,z$) correspond to Pauli matrices
\begin{equation} 
\sigma_x = \begin{pmatrix} 0 &1\\1&0 \end{pmatrix},
\end{equation}
\begin{equation} 
\sigma_y = \begin{pmatrix} 0 &-i\\i&0 \end{pmatrix},
\end{equation}
\begin{equation} 
\sigma_z = \begin{pmatrix} 1 &0\\0&-1 \end{pmatrix}.
\end{equation}
A $R_y(\theta)$ rotation is always added before the measurements, which can be interpreted as optimizing the observable which is measured. Thus, the final unitary can be written as 
\begin{align}
\label{eq:U}
    U(\Theta) &=  \left(\bigotimes_{j=1}^N R_y^j((\vec{\Theta}_L)_j)\right)  \prod_{i=L-1}^{0} U_i(\vec{\Theta}_i)\\ 
   = &  \left(\bigotimes_{j=1}^N R_y^j((\vec{\Theta}_L)_j)\right) \prod_{i=L-1}^{0} \Biggl[\left(\prod_{k=1}^{N-1} \text{CX}_{k,k+1}\right)\cdot \\
   \times & \left(\bigotimes_{j=1}^N R_x^j((\vec{\Theta}_i)_{N+j})\right)  \left(\bigotimes_{j=1}^{N} R_y^j((\vec{\Theta}_i)_{j})\right) \Biggr]
\end{align}
where $R_y^j(\theta)$ $(R_y^j(\theta))$ is a rotation of the $j$th qubit around the $y (x)$ axis with angle $\theta$ \cite{nielsen_chuang_2010} and CX$_{k,j}$ is a CNOT \cite{nielsen_chuang_2010} gate between qubits $k$ and $j$.

\subsection{Correlated features}
\label{3DBM}
\begin{figure}
    \centering
    \includegraphics[scale=0.39]{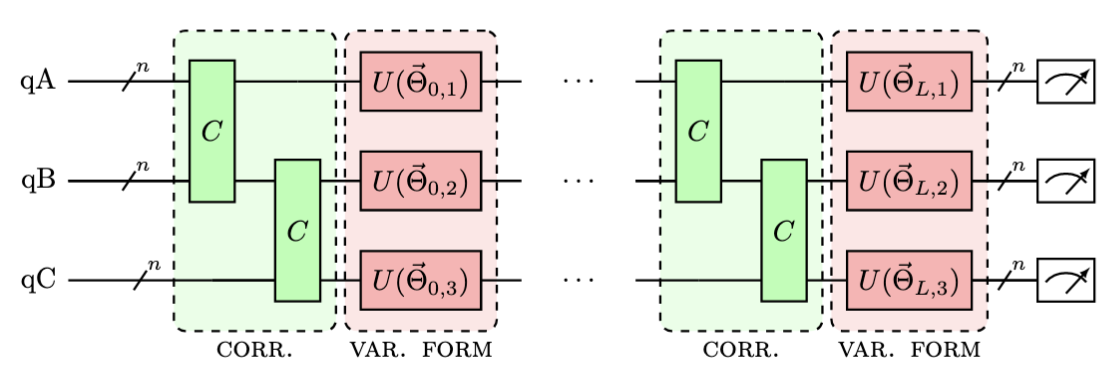} 
    \caption{QCBM for a multivariate probability distribution. The fixed $C$ gates create entanglement between the registers while the trainable $U(\vec{\Theta}_{i,j})$ block learns the distributions.}
    \label{fig:corr_circuit}
 \end{figure}

\begin{figure}
    \centering
    \includegraphics[scale=0.5]{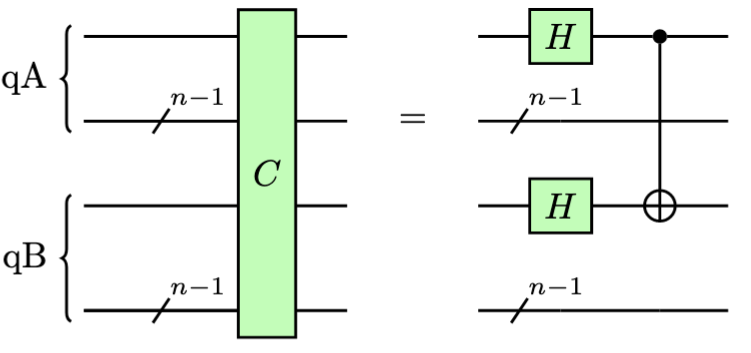}
    \caption{Correlation gates $C$ between two quantum registers $qA$ and $qB$,  with $n$ qubits each.}
    \label{fig:G2}
\end{figure}

A simple way to extend the above\hyp defined QCBM to generate $D$ correlated features is to use different registers ($qA$, $qB$, $qC$, etc) for each of them, as proposed in Ref. \cite{Joint_distribution}. We will therefore associate each feature with a quantum register of size $n$, where the total number of qubits needed is $Dn$. In this scenario, a fixed unitary $C$ entangles the registers while local operators $U_{i,j}(\vec{\Theta}_{i,j})$ learn the individual distributions, as shown in \figref{fig:corr_circuit}. The index $i$ refers to the layer, and $j$ refers to the register. While the local operators are trainable, the correlation gates $C$ do not contain any free parameters. This ensures that the number of parameters is kept to a minimum, considering that rotations are already included in the local operators. The operator $C$ is built with the following two-qubit block
\begin{equation}
   ( H\otimes H) \cdot \text{CX}_{qA_i,qB_i},
\end{equation}
which is used to entangle the $i$th qubit of the $qA$ and $qB$ register. Here, $H$ refers to the Hadamard gate \cite{nielsen_chuang_2010}. We consider different variations of this setup:
\begin{enumerate}
   
    \item We vary the registers are entangled together, e.g. in a \emph{linear} way, where each register is connected to the next one, or in a \emph{full} way, where every register is connected to all the others. \\
    \item We vary the number of qubits in each register which are acted upon, e.g., only the first pair (denoted by $j = 1$), or all of them ($j=n$, denoted by "all"). More precisely, the $qA_i$ qubit is connected to the $qB_i$ one for $i \in \{ 1,..., j\}.$\\
    \item Finally, we can choose the block to be 
    \begin{equation}
        ( H\otimes \mathbb{1}) \cdot \text{CX}_{qA_i,qB_i},
    \end{equation}
    which constructs a Bell state, when two registers are involved or, \begin{equation}
        ( H\otimes \mathbb{1} \otimes \mathbb{1}) \cdot \text{CX}_{qA_i,qB_i} \text{CX}_{qB_i,qC_i},
    \end{equation} which constructs a Greenberger\hyp Horne\hyp Zeilinger state when there are three registers. This choice will be denoted by the label "Bell".
\end{enumerate}
This gives eight possibilities in total, and they were all tested for the considered use case. A comparison will be shown in Sec. \ref{res:3DBM}. However, we can already mention that an easier choice, such as (\emph{linear}, 1) or (\emph{linear}, 1, Bell), with a reduced connectivity and number of gates usually leads to higher performance, in terms of both the marginal distribution and correlations. Moreover, while the more expensive option (\emph{linear}, all) achieves a smaller loss on the simulator, it failed on the hardware since the execution time exceeded the coherence time of the device, producing uniform distributions from maximally mixed states. We therefore advocate for the use of the minimal (\emph{linear}, 1) block, which can be constructed using the circuit depicted in \figref{fig:G2}.


\begin{figure}
    \centering
    \includegraphics[scale=0.5]{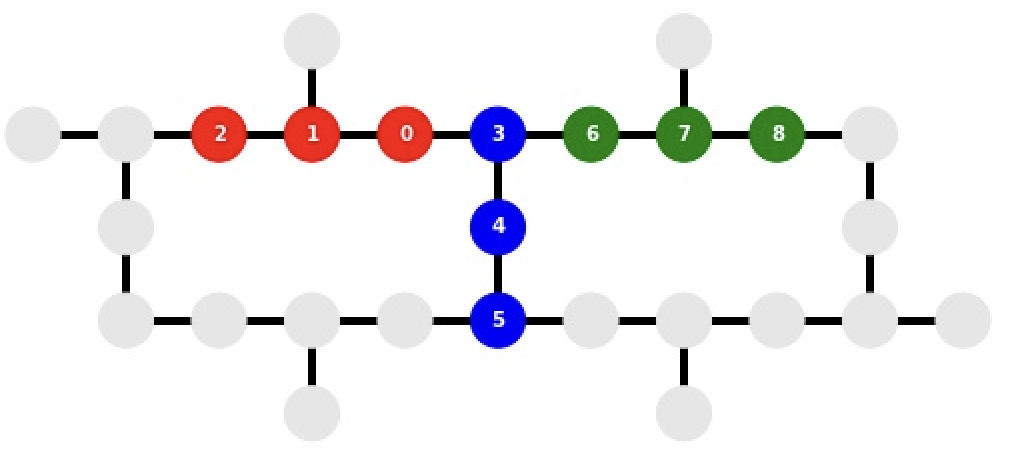}
    \caption{Mapping of the multidimensional QCBM onto a 27\hyp qubit IBM quantum chip. The different colors refer to different registers and the ordering follows the convention of QISKIT.}
    \label{topology}
\end{figure}

Moreover, it is possible, using the circuit in  ~\figref{fig:U_ha} for $U_{i,j}(\vec{\Theta}_{i,j})$, for all $i$ and $j$, to map the multivariate Born machine from ~\figref{fig:corr_circuit}, with the (\emph{linear}, 1) choice to an IBM Quantum chip without using any SWAP gates. This is due to the linear connectivity of all the components and the \emph{T} topology of the device. \figref{topology} shows a possible way to do so onto a 27\hyp qubit architecture using three registers ($qA$ in red, $qB$ in blue, and $qC$ in green) with $n=3$ qubits each. The elimination of SWAP gates diminishes the number of errors made on the quantum devices by reducing the number of two\hyp qubit gates and depth. 

\subsection{Conditional Born machine}
\label{CBM}
Conditional generative models, such as conditional generative adversarial networks \cite{CGAN}  produce samples $x$ according to some conditions $y$. This task is more challenging since $p(x|y)$ has to be captured, instead of only $p(x)$, where $p(x)$ is the probability of event $x$ happening. The flexibility of conditional generative models compared to MC simulations is advantageous in terms of the computational and time resources needed to generate complex events. For instance, in the MC simulations used in this work, the initial energy of the incoming muon has to be fixed, while it is variable in ML-based techniques. This is a strong argument in favor of machine learning, both classical and quantum, for data generation in HEP. Hence, while the MC computations have to be performed for all the different values of the conditioning variable, machine learning models can learn from a reduced training set, and interpolate, or even extrapolate, considerably reducing the time consumption needed for MC simulations.

\begin{figure}
    \centering
    \includegraphics[scale=0.45]{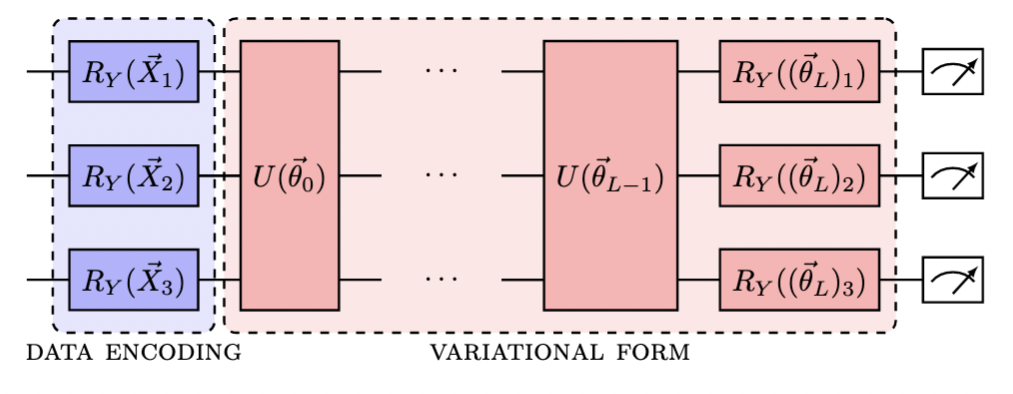}
    \caption{Conditional Born machine. The data dependent block (blue) acts as a feature map while the trainable gates (red) learn the distribution.}
    \label{fig:CBorn}
\end{figure}

Condition $y$ in MFC events is the energy $E_{\text{in}}$ of the incoming muon.  Different experimental values for $E_{\text{in}}$ are considered, ranging from 50 to 200 GeV in steps of 25 GeV. The conditional QCBM's goal is to generate the correct distributions when given access to the incoming muon's energy. In practice, $y=E_{\text{in}}$ is first scaled between [0,1], is transformed with  the function arcsine, as used in \cite{QNN_FF}, and is then encoded into the QCBM via repeated $y$ rotations on all qubits,  as shown in \figref{fig:CBorn}.
The preprocessing ensures that the data are in the right range to be interpreted as an angle. Overall, the model consists of a feature map $\Phi(\vec{X})$ which encodes the data $\vec{X}$ and trainable gates that learn the probability distribution, and can be written as 
\begin{align}
    U(\vec{X},\Theta) =U(\Theta) \Phi(\vec{X}),
\end{align}
with $U(\Theta)$ as in Eq. \ref{eq:U}, $\Phi(\vec{X})$ being the data\hyp encoding feature map 
\begin{align}
    \Phi(\vec{X}) =  \bigotimes_{j=1}^N R_y(\vec{X}_j) 
\end{align} and 
\begin{equation}
    \vec{X}_i = \arcsin{(\text{minmax}[E_{\text{in}},0,1)]}.
\end{equation}
Here, minmax$(Y,a,b)$ scales the data set $Y$ between the values $a$ and $b$.

More complex feature maps \cite{feature_map}, and data reuploading \cite{Schuld_Fourier} strategies were tested but did not show any improvements.

\section{Training Strategy}
\label{training}
\subsection{Optimization}
The QCBM is trained using a two\hyp sample test \cite{MMD}, with a Gaussian kernel 
\begin{equation}
    K(x,y) = \text{exp}(-\frac{(x-y)^2}{2\sigma}),
\end{equation} by comparing the distance between two samples $x$ and $y$ in the kernel feature space. Concretely, the maximum mean discrepancy (MMD) \cite{MMD} loss function
\begin{align}
\label{MMD}
    \mathcal{L} = \underset{x\sim p_\theta, y\sim p_\theta}{\mathbb{E}[K(x,y)]} -2\underset{x\sim p_\theta, y\sim \pi}{\mathbb{E}[K(x,y)]} + \underset{x\sim \pi, y\sim \pi}{\mathbb{E}[K(x,y)]}
\end{align}
is used, with bandwidth 
\begin{equation}
     \sigma \in [0.01,0.1,1,10,100].
\end{equation}
In this way, the differences of all the moments between the target and model probability distributions are efficiently compared at different scales. The advantages of the MMD include its metric properties and the training stability it provides, making it a suitable option in the NISQ era.

The gradient can be computed \cite{Born_machine_Liu} using the parameter\hyp shift rule \cite{quantum_grad} as 
\begin{align}
    \frac{\partial \mathcal{L}}{\partial \theta_i} &=  \underset{x\sim p_{\theta^+}, y\sim p_\theta}{\mathbb{E}[K(x,y)]} - \underset{x\sim p_{\theta^-}, y\sim p_\theta}{\mathbb{E}[K(x,y)]} \nonumber \\
    & -\underset{x\sim p_{\theta^+}, y\sim \pi}{\mathbb{E}[K(x,y)]} + \underset{x\sim p_{\theta^-}, y\sim \pi}{\mathbb{E}[K(x,y)]} ,
\end{align}
where $p_{\theta^{\pm}}$ are QCBMs with parameters $\theta^{\pm} = (\theta \pm  \pi/2) \hat{e}_i$ with $\hat{e}_i$ being the $i$th unit basis vector in the parameter space, i.e. $(\hat{e}_i)_j = \delta_{ij}$.

Alternatively, the simultaneous perturbation stochastic approximation (SPSA) \cite{SPSA} algorithm is also considered to optimize the QCBM in a gradient\hyp free fashion. SPSA efficiently approximates the gradient with two sampling steps by perturbing the parameters in all directions simultaneously. While the convergence is slower than using the exact gradient, fewer circuit evaluations are needed for each epoch. Moreover, the stochastic nature of SPSA makes it more resilient to hardware and statistical noise. It was observed during the simulations that the gradient-based algorithm outperforms SPSA, as SPSA sometimes gets trapped in local minima. However the gradient-based algorithm is more resource intensive than SPSA, and in this regime, SPSA is often preferred as it is better suited for quantum hardware.

Therefore, a mixed training scheme is used, where the models are first trained on (noisy) simulators using the adaptive moment estimator (ADAM) optimizer \cite{adam} and then fine\hyp tuned for a few epochs on quantum hardware using SPSA. A readout\hyp error\hyp mitigation scheme \cite{MEM} is used on the measurements. Details about the implementation, training, and resources can be found in the Appendix. 

\subsection{Classical baseline}

Classical generative models trained using the MMD loss function (GMMD models) \cite{GMMD,training_GMMD} are used as a baseline. They are trained on continuous data since the performance is usually higher than for discrete samples. The considered model is a simple fully connected neural network with only a few thousand parameters. It is highly probable that higher\hyp performing models can be designed with some care. For instance, \cite{MFC_GAN} shows a solution to the same problem using GAN. The goal of the classical baseline at this stage is to give an indication of the current level of deployability of quantum machine learning models and not to predict quantum advantage.  


\section{Results}
\label{res}

\subsection{Experimental device}
\label{hardware_description}
The quantum devices (IBMQ Montreal and IBMQ Mumbai \cite{Jurcevic_2021}) used in this work consist of two 27\hyp fixed\hyp frequency transmon qubits, with fundamental transition frequencies of approximately 5 GHz and anharmonicities of $-340$ MHz. Their topology is displayed in \figref{topology}. Microwave pulses are used for single\hyp qubit gates and cross\hyp resonance interaction \cite{cross-resonance} is used for two\hyp qubit gates. The experiments took place over 3 hours each, without intermediate calibration. The median qubit lifetime $T_1$ of the qubits are 121 and 129 $\mu s$, the median coherence time $T_2$ are 90 and 135 $\mu s$ and the median readout errors are 0.029 and 0.021 for the two devices, respectively. The qubits, which are used in the experiments, are chosen such that the total CNOT and readout error are minimized. The CNOT error varies between 0.006 and 0.02, depending on the specific connection.

\subsection{One dimensional distribution}
\begin{figure}
    \centering
    \includegraphics[scale=0.7]{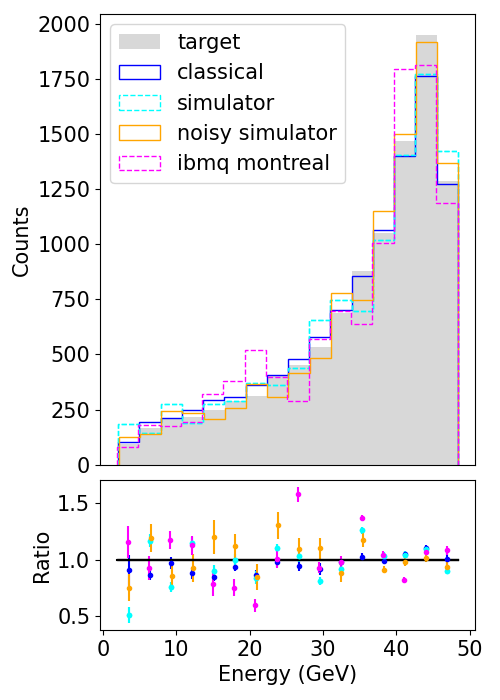}
    \caption{The outgoing muon's energy at 50 GeV. Comparison of QCBMs run on a perfect simulator (dashed cyan), noisy simulator (orange) and IBMQ Montreal (dashed pink) with the classical GMMD (blue).}
    \label{fig:1d_results}
\end{figure}

As a first demonstration, the QCBM is trained on a one\hyp dimensional distribution: the energy of the outgoing muon discretized on $2^4= 16$ bins. The QCBM is built with one repetition of $R_Y(\theta)$ and $R_X(\theta)$ gates on all qubits  and $(R_{zz})_{i,j}(\theta) = \exp{(-i\theta \sigma_z^i \otimes \sigma_z^j)}$ interaction on qubit $i$ and $j$, using a full entanglement scheme. The unitary can thus be written as 
\begin{align}
U(\Theta) &= \left(\bigotimes_{i=1}^{4} 
R_y((\vec{\Theta}_1)_{i})\right) \\
& \prod_{i=1}^{3}\prod_{j=i+1}^4 (R_{zz})_{i,j}((\vec{\Theta}_0)_{(8+4*i+j}) \\
&\times \left(\bigotimes_{i=1}^{4} R_x((\vec{\Theta}_0)_{4+i})
R_y((\vec{\Theta}_0)_{i})\right),
\end{align}
and has 18 parameters. This particular structure has been found to be best suited for the current situation via trial and error. In particular, empirical evidence suggests that this circuit is better suited than the one proposed in \figref{fig:U_ha}. The small number of two\hyp qubit gates enables the use of real quantum hardware without severe complications due to the noise. Results obtained with an ideal simulator, noisy simulator, superconducting circuits (IBMQ Montreal) and classical GMMD are shown in \figref{fig:1d_results}. The histograms display the number of generated events and the ratios with the data set as a function of energy (GeV), with error bars corresponding to one standard deviation from 10 sampling processes.
The GMMD is chosen to be a neural network with four hidden layers of size $[64, 128, 64, 16]$, each with a sigmoid activation function, and a latent space of dimension 15.

The total variance (TV) $\in [0,1]$ with sample set $\Omega$,
\begin{equation}
    V_T(p,\pi)= \frac{1}{2}\sum_{x\in \Omega}|p(x)-\pi(x)|
\end{equation}
is used as a comparison metric and results are shown in Table \ref{tab:1d}.\\
\begin{table}
    \centering
    \begin{tabular}{c|c}
    \textbf{Backend} & \textbf{TV}\\
    \hline 
      Simulator   & 0.055\\
       Noisy simulator & 0.043\\ 
       IBMQ Montreal & 0.074 \\ 
        GMMD & 0.028 \\ 
       \hline  
    \end{tabular}
    \vspace{0.11cm}
    \captionsetup{aboveskip=10pt}
    \caption{Total variance (TV) for the one dimensional distribution on different backends.}
    \label{tab:1d}
\end{table}
Although slightly outperformed by the GMMD, the QCBM can still be competitive even with its small number of parameters. The noise does not negatively contribute to the performance, as emphasized by the noisy simulations and quantum hardware results.

\subsection{Multivariate distribution}
\label{res:3DBM}
\begin{figure}
    \centering
    \includegraphics[scale=0.6]{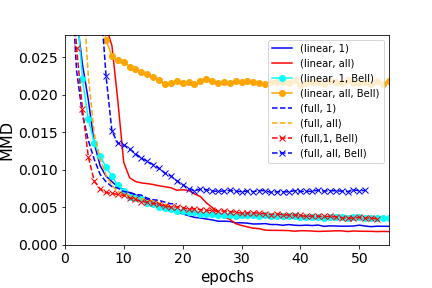}
    \caption{Validation MMD loss values during  training for the eight different architecture choices for the correlation block considered in the present paper.}
    \label{loss_3d_qcbm}
\end{figure}

As a next step, we consider a multivariate distribution, namely, the energy, transverse momentum, and pseudorapidity of the outgoing muon with an incoming energy of 125 GeV, using $2^3=8$ bins. The QCBM is designed with four repetitions of entangling $C$ and a local $U_i(\vec{\Theta}_{i,j})$ block, as seen in \figref{fig:corr_circuit}. The former creates an entangling state, while the latter consists of $R_y(\theta)$ and CNOT interaction with linear connectivity. The model therefore has  $45$ parameters. We first assess the performance of the different choices for the correlation block described in Sec. \ref{3DBM}  by showing the validation MMD loss values during the training in \figref{loss_3d_qcbm}. We observe that the (\emph{linear}, 1) and (\emph{linear}, all) blocks perform similarly and are the best choices for the present task. It is not surprising that the best results are obtained with the long\hyp range interaction provided by (\emph{linear, all}), as this outcome was also reported in \cite{Born_machine_Coyle_2021,Joint_distribution}. However, the performance obtained on the quantum hardware is improved by using the (\emph{linear}, 1) block since it contains fewer CNOT gates. We will therefore choose this architecture for the rest of the paper. 

\begin{figure*}
        \centering
        \begin{subfigure}[b]{0.32\textwidth}
            \centering
            \includegraphics[width=\textwidth]{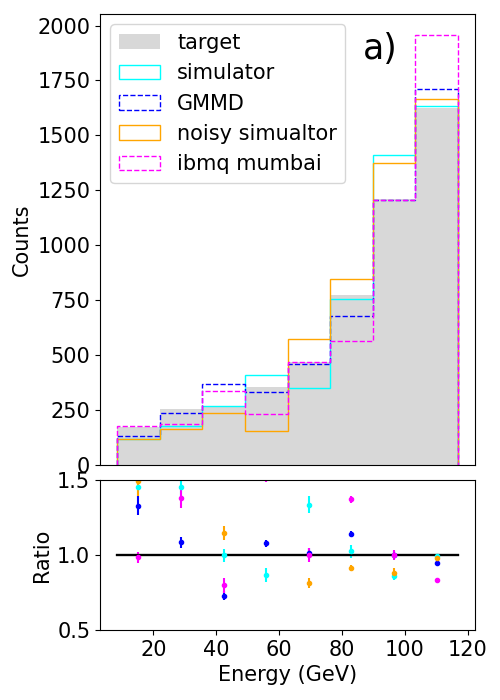}
                
        \end{subfigure}
        \hfill
        \begin{subfigure}[b]{0.32\textwidth}  
            \centering 
            \includegraphics[width=\textwidth]{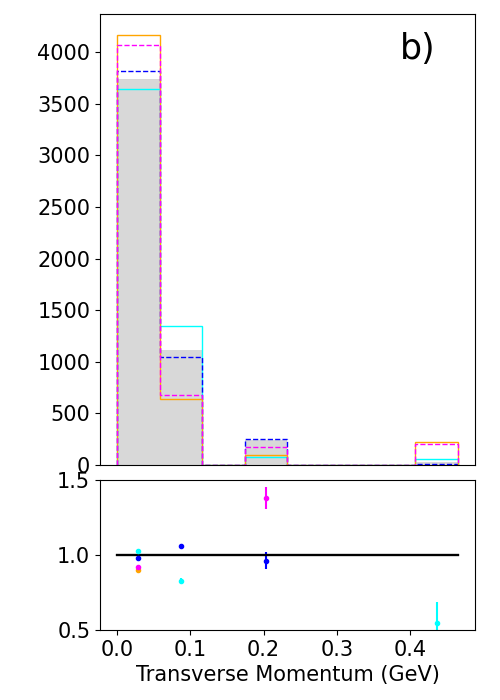}
              
        \end{subfigure}
        \hfill
        \begin{subfigure}[b]{0.32\textwidth}   
            \centering 
            \includegraphics[width=\textwidth]{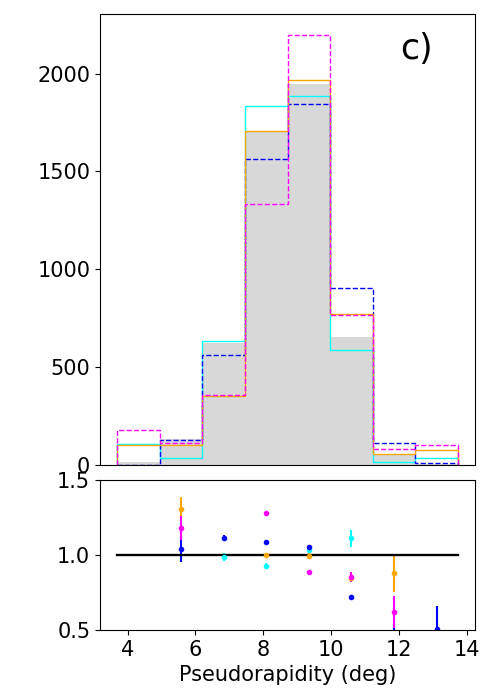}
               
            \label{fig:mean and std of net34}
        \end{subfigure}
       
\caption{ (a) The outgoing muon's energy, (b) transverse momentum and (c) pseudorapidity at 125 GeV, generated by the multidimensional QCBM on a perfect simulator (cyan), noisy simulator (orange) and IBMQ Mumbai (dashed pink) compared to a classical GMMD (dashed blue).}
\label{fig:multi_dim}
    \end{figure*}

 The results for the simulator, noisy simulator, IBMQ Mumbai, and GMMD are shown in \figref{fig:multi_dim} and the total variance for the marginal distributions are presented in Table \ref{tab:3d}. The GMMD is constructed similarly to that above but with three hidden layers of size $[128,256,128]$. Even if the GMMD achieves the best accuracy, the QCBM is still competitive despite its small number of learned parameters or the presence of noise.




\begin{table}
    \centering
    \begin{tabular}{c|c|c|c}
    \textbf{Backend} & $V_T(E)$ & $V_T(p_t)$ & $V_T(\eta)$\\
    \hline 
      Simulator  &0.055&0.05 & 0.052\\

       Noisy simulator &0.075&0.12& 0.06 \\ 
       IBMQ Mumbai & 0.078&0.097&0.13 \\ 
        GMMD &0.036&0.017& 0.063 \\ 
       \hline  
    \end{tabular}
    \vspace{0.11cm}
    \captionsetup{aboveskip=10pt}
    \caption{Total variance for the individual multivariate distributions on different backends.}
    \label{tab:3d}
\end{table}

An important factor for the performance of generative models is their ability to learn the correlations between the variables, which is not reflected in the total variance. To this point, the correlations in the target dataset are compared to those in the generated datasets. The correlation matrices are computed with the Pearson product-moment methods 
\begin{equation}
    R_{ij} = \frac{C_{ij}}{\sqrt{C_{ii}\cdot C_{jj}}},
\end{equation}
where $C$ is the covariance matrix. The correlations in the target dataset (ground truth) are shown in Table \ref{correlation_target} and those for the generated samples are in Table \ref{corr_samples}.

\begin{table}[]
\Large
\begin{tabular}{c|c|c|c|}
\cline{2-4}
                         & $E$                                              & $p_t$                                             & $\eta$ \\ \hline
\multicolumn{1}{|c|}{$E$}  & \multicolumn{1}{c|}{\cellcolor[HTML]{C0C0C0}-} & 0.43                                           & 0.89   \\ \hline
\multicolumn{1}{|c|}{$p_t$} & 0.43                                           & \multicolumn{1}{c|}{\cellcolor[HTML]{C0C0C0}-} & 0.61   \\ \hline
\multicolumn{1}{|c|}{$\eta$} & 0.89 & 0.61 & \multicolumn{1}{c|}{\cellcolor[HTML]{C0C0C0}{\color[HTML]{000000} -}} \\ \hline
\end{tabular}

\caption{MC correlations (ground truth)}
\label{correlation_target}
\end{table}

\begin{table*}[]
\begin{tabular}{l|cccc|cccc|cccc|l}
\cline{2-13}
 &
  \multicolumn{4}{c|}{$E$} &
  \multicolumn{4}{c|}{$p_t$} &
  \multicolumn{4}{c|}{$\eta$} &
   \\ \cline{2-13}
 &
  \multicolumn{1}{l|}{sim} &
  \multicolumn{1}{l|}{noisy} &
  \multicolumn{1}{l|}{IBMQ} &
  \multicolumn{1}{l|}{GMMD} &
  \multicolumn{1}{l|}{sim.} &
  \multicolumn{1}{l|}{noisy} &
  \multicolumn{1}{l|}{IBMQ} &
  \multicolumn{1}{l|}{GMMD} &
  \multicolumn{1}{l|}{sim.} &
  \multicolumn{1}{l|}{noisy} &
  \multicolumn{1}{l|}{IBMQ} &
  \multicolumn{1}{l|}{GMMD} &
   \\ \cline{1-13}
\multicolumn{1}{|c|}{$E$} &
  \multicolumn{4}{c|}{\cellcolor[HTML]{C0C0C0}-} &
  \multicolumn{1}{c|}{\cellcolor[HTML]{38FFF8}0.6} &
  \multicolumn{1}{c|}{\cellcolor[HTML]{FFC702}0.3} &
  \multicolumn{1}{c|}{\cellcolor[HTML]{EC71FE}0.29} &
  \cellcolor[HTML]{3531FF}{\color[HTML]{FFFFFF} 0.44} &
  \multicolumn{1}{c|}{\cellcolor[HTML]{38FFF8}0.9} &
  \multicolumn{1}{c|}{\cellcolor[HTML]{FFCB2F}0.87} &
  \multicolumn{1}{c|}{\cellcolor[HTML]{EC71FE}0.88} &
  \cellcolor[HTML]{3531FF}{\color[HTML]{FFFFFF} 0.91} &
   \\ \cline{1-13}
\multicolumn{1}{|c|}{$p_t$} &
  \multicolumn{1}{c|}{\cellcolor[HTML]{38FFF8}0.6} &
  \multicolumn{1}{c|}{\cellcolor[HTML]{FFC702}0.3} &
  \multicolumn{1}{c|}{\cellcolor[HTML]{EC71FE}0.29} &
  \cellcolor[HTML]{3531FF}{\color[HTML]{FFFFFF} 0.44} &
  \multicolumn{4}{c|}{\cellcolor[HTML]{C0C0C0}-} &
  \multicolumn{1}{c|}{\cellcolor[HTML]{38FFF8}0.79} &
  \multicolumn{1}{c|}{\cellcolor[HTML]{FFCB2F}0.58} &
  \multicolumn{1}{c|}{\cellcolor[HTML]{EC71FE}0.6} &
  \cellcolor[HTML]{3531FF}{\color[HTML]{FFFFFF} 0.61} &
   \\ \cline{1-13}
\multicolumn{1}{|l|}{$\eta$} &
  \multicolumn{1}{c|}{\cellcolor[HTML]{38FFF8}0.9} &
  \multicolumn{1}{c|}{\cellcolor[HTML]{FFC702}0.87} &
  \multicolumn{1}{c|}{\cellcolor[HTML]{EC71FE}0.88} &
  \cellcolor[HTML]{3531FF}{\color[HTML]{FFFFFF} 0.91} &
  \multicolumn{1}{c|}{\cellcolor[HTML]{38FFF8}0.79} &
  \multicolumn{1}{c|}{\cellcolor[HTML]{FFC702}0.58} &
  \multicolumn{1}{c|}{\cellcolor[HTML]{EC71FE}0.6} &
  \cellcolor[HTML]{3531FF}{\color[HTML]{FFFFFF} 0.61} &
  \multicolumn{4}{c|}{\cellcolor[HTML]{C0C0C0}-} &
   \\ \cline{1-13}
\end{tabular}

\caption{Correlation in the generated samples obtained on the simulator (light blue), noisy simulator (orange), IBMQ Mumbai (pink) and with the classical GMMD (dark blue).}
\label{corr_samples}
\end{table*}

We observe that the QCBMs trained on the different backends are able to capture the correct correlations, even if the classical GMMD is better. It is noteworthy that the samples generated by the quantum chip are closer to the ground truth than the simulated ones, suggesting that generative modeling is a promising task for NISQ devices.

\subsection{Conditional distribution}
\begin{figure*}
        \centering
        \begin{subfigure}[b]{0.32\textwidth}
            \centering
            \includegraphics[width=\textwidth]{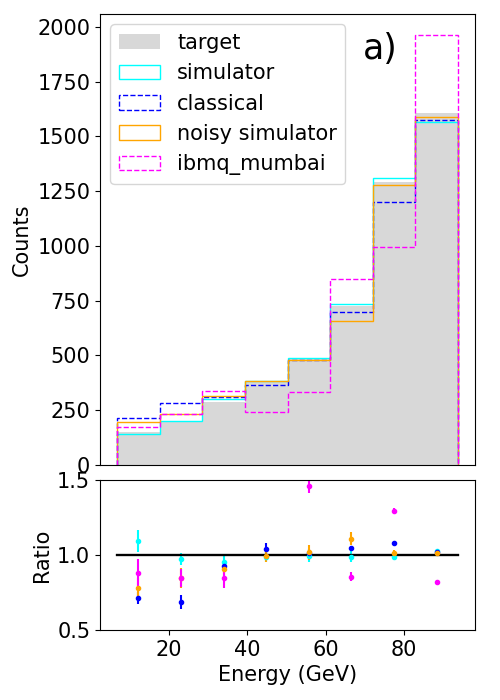}
               
        \end{subfigure}
        \hfill
        \begin{subfigure}[b]{0.32\textwidth}  
            \centering 
            \includegraphics[width=\textwidth]{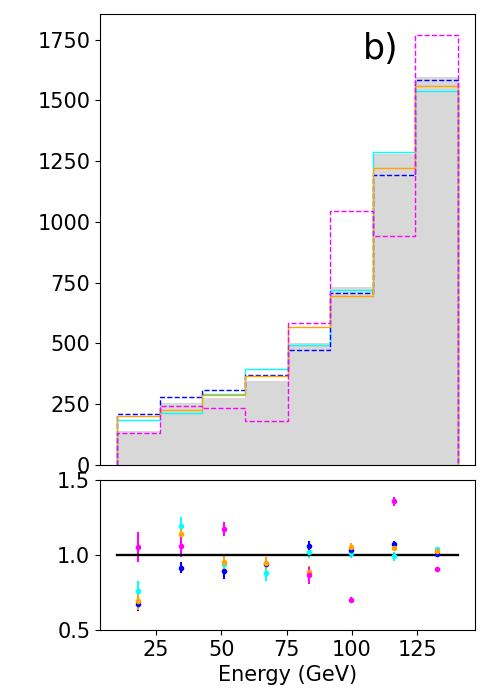}
                
        \end{subfigure}
        \hfill
        \begin{subfigure}[b]{0.32\textwidth}   
            \centering 
            \includegraphics[width=\textwidth]{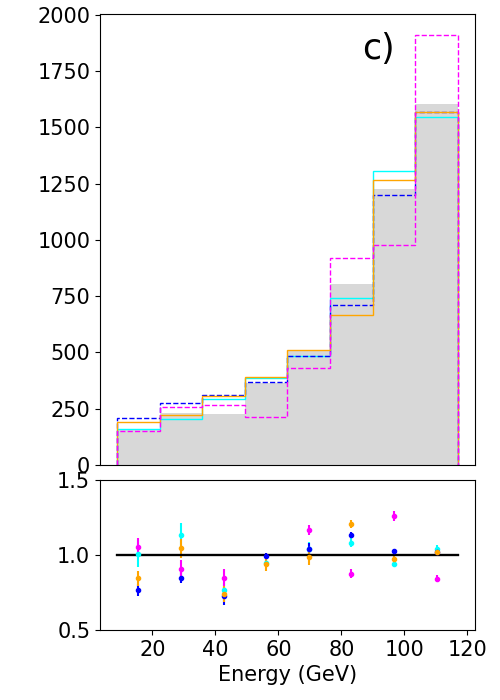}
               
            \label{fig:mean and std of net34}
        \end{subfigure}

\caption{The outgoing muon’s energy with an initial energy of (a) 100 GeV (train), (b) 150 GeV (train), and (c) 125 GeV (test) generated by the QCBM on a perfect simulator (cyan), noisy simulator (orange), and IBMQ Mumbai (dashed pink) compared to a classical GMMD (dashed blue). The models are tested on samples with an energy of 125 GeV while they were trained on samples with the remaining energies.}
\label{fig:cborn}
\end{figure*}

Finally, we consider conditional QCBM for the generation of MC events conditioned on the initial muon's energy, which is encoded in the QCBM via parametrized\hyp rotations. The QCBM, as outlined in \figref{fig:CBorn} contains four repetitions of $U_i(\vec{\Theta}_i)$, with $R_y(\theta)$ and $R_x(\theta)$ rotations (with different parameters) and CNOT interaction in a linear fashion as depicted in \figref{fig:U_ha}, while the GMMD has two hidden layers of size $[8, 8]$. The thus QCBM contains 27 trainable parameters. The training is performed on the whole dataset except at 125 GeV, which is left to test the interpolation capabilities of the models. Results are shown in \figref{fig:cborn}, and the values of the total variance are reported in Table \ref{tab:cborn}. All models achieve good performance for the interpolation. The results on the quantum hardware could be slightly improved for some histogram binned values. However, the performance is similar on training and testing energy bins, suggesting that the QCBM can interpolate but is strongly affected by the noise.

\begin{table}
    \centering
    \begin{tabular}{c|c|c|c}
    \textbf{Backend} & $V_T(100)$& $V_T(150)$ & $V_T(125)$\\
    \hline 
      Simulator  &0.033&0.016 & 0.033\\
       Noisy simulator &0.067& 0.046&0.035\\ 
       IBMQ Mumbai & 0.15&0.13&0.094 \\ 
    GMMD &0.016&0.032& 0.034 \\
     
       \hline  
    \end{tabular}
    \vspace{0.11cm}
   \captionsetup{aboveskip=10pt}
    \caption{Total variance for the conditional distribution on different backends.}
    \label{tab:cborn}
\end{table}

\section{Discussion}
\label{discussion}
The results presented in the Sec. \ref{res} suggest that QCBMs can reproduce the marginal distribution, as well as the correlations, from MC simulations. Even if a higher performance can easily be obtained with classical neural networks, it is important to underline that QCBMs generally operate with very few parameters for similar performance. This suggests that QCBMs are more expressive than classical neural networks, as outlined by \textcite{abbas_power_2021}, and outperform them in the under\hyp parametrized regime. It remains an open question whether QCBMs enjoy a higher performance in the overparameterized regime, as is the case for classical models \cite{deep_DD}. We note that this question was already explored for quantum neural networks by \textcite{overparametrization}.

Moreover, the presence of noise does not seem to be an obstacle to the training of QCBMs. It is noteworthy that the results obtained on quantum hardware are close to that obtained on the simulator, which suggests sufficient device quality for this task and an ability to deal with incoherent noise. The hardware results are slightly worse in the conditional case which can be explained by the reduced number of epochs performed on the quantum hardware. The loop over the training energy bins increases the resources needed for one epoch, and thus reduces the number of epochs performed.

These observations suggest that the noise is assimilated during the training, underlining the importance of using actual quantum hardware. This supports the findings of \textcite{Rhem_noise}, which empirically found that quantum generative adversarial networks can be efficiently trained on quantum hardware if the readout noise is smaller than 0.1. Thus, QCBMs seem to be an appealing application for NISQ devices.

Barren plateaus (BP) are large portions of the training landscape where the loss function's gradient variance vanishes. As shown in \cite{Barren_platea_McClean}, BPs appear exponentially fast in the depth and number of qubits for generic quantum circuits, which makes the training of large\hyp scale quantum variational algorithms generally difficult. Solutions to this issue, such as quantum convolutional neural networks \cite{QCNN} and local loss functions \cite{local_loss} are not applicable in this case since measurements on all qubits are needed.

However, BPs have not been observed in this work, and were not reported in similar studies \cite{Joint_distribution,QCBM_MC} either. This can be explained to some extent by the relative shallowness of the circuits used. Nevertheless, difficulties are observed durings training on hardware and noisy simulators, which could be an effect of noise\hyp induced BPs \cite{noise_BP}. A more significant number of epochs was needed to mitigate this effect. Increasing the number of qubits to five or six, for the one\hyp dimensional case, causes the ratio between the generated and target samples for each bin to deteriorate, even if the loss function converged after a hundred epochs. The same problem appeared when increasing the number of features in the multivariate case and mixing multiple features with the conditioning. Since the gradient never vanished at the beginning of the training, BPs are probably not the most critical issue, on the other hand, the MMD may not be the most suitable loss function for large\hyp scale QCBMs.

Alternative training strategies were proposed in  \cite{Born_machine_Coyle_2021,QGAN_Zoufal}, with optimal transport and an adversarial training strategy, respectively. Hence, empirical evidence suggests that the strong theoretical properties of the MMD loss function are not met in practice, as outlined by some benchmarks \cite{benchmark}. Hence, the performances of GMMD and GAN are similar for simple problems but the latter is superior for complex tasks. \textcite{MMD_GAN} proposed an adversarial strategy to optimize the kernel as an efficient way to improve the performance of GMMD models.

\section{Conclusion}
The present paper presented the application and further development of a quantum circuit Born machine to generate Monte Carlo events in HEP, specifically muon force carriers. An efficient way to generate multivariate distributions, requiring only linear connectivity and thus being suitable for NISQ devices, was proposed. Additionally, the present paper took a step towards generating conditional probability distributions with quantum circuit Born machines. Numerical evidence demonstrated that QCBMs can efficiently generate joint and conditional distributions with the correct correlations. Finally, the experiments were run successfully on quantum hardware, hinting that QML algorithms can mitigate the effect of the noise during the training. Quantum generative models are consequently appealing for NISQ devices since they can manage noisy qubits without the need of for expensive error\hyp mitigation techniques. QCBMs also have the advantage of needing a small number of parameters while still being competitive. 

While having strong potential in generative modeling, QCBMs still need some improvement to handle a more refined binning and multivariate distribution of higher dimensions. Additionally, it would be interesting to consider conditional distributions which are more sensitive to the conditioning variable, and they will be the focus of future work.

\section*{Acknowledgement}
The authors thank S. Y. Chang, D. Pasquali, T. Ramazyan, and F. Rehm for valuable and stimulating discussions about the present paper.

This work was supported by the CERN Quantum Technology Initiative. Simulations were performed on the University of Geneva's Yggdrasil HPC cluster. Access to the IBM Quantum Services was obtained through the IBM Quantum Hub at CERN. The views expressed are those of the authors and do not reflect the official policy or position of IBM or the IBM~Q team. 

\bibliography{bibliography}

\appendix

\section{Implementation}\label{ressources}
The noiseless simulations are performed with PENNYLANE \cite{pennylane} powered by \href{https://pennylane.ai/qml/demos/tutorial_jax_transformations.html}{JAX} \cite{jax2018github}, which enables an efficient gradient computation via vectorization and just\hyp in\hyp time compilation. The noisy simulations are performed using a fake backend tuned to the real quantum hardware, provided by QISKIT \cite{Qiskit}. The fake backend has the same characteristics on average as the real backend described in Sec. \ref{hardware_description}. The training is performed in batches composed of 512 events each, and one epoch is composed of 10 batches. The learning rate is initially set to 0.01 and is halved every 20 epochs.
The resources needed to produce the presented results are presented in Table \ref{tab:ressources}, which shows the number of parameters, the time needed for a forward pass and a backward pass, and the number of epochs until convergence for all the quantum models trained on the simulator. Each epoch is composed of 10 batches, except the conditional model which has 10 batches per training energy bin (i.e., six). Each batch contained 512 samples. Simulations were run on a single CPU on the University of Geneva's Yggdrasil HPC cluster.

\begin{table}[htp]
    \centering
    \begin{tabular}{c|cccc}
    Model & Param. & Forward  & Backward & Epochs\\
     &  &  pass (s) &  pass (s) & \\
    
    \hline 
      One\hyp dimensional & 18 &1.2 &3.9 &70 \\
      Multivariate & 45 &1.9 &9.4  &100\\
      Conditional & 27&1.5 &4.5  & 30 \\
       \hline 
    \end{tabular}
    \caption{Number of parameters, time needed for a forward and backward pass and number of epochs until convergence for the three quantum models trained on a simulator.}
    \label{tab:ressources}
\end{table}

\end{document}